\documentclass[a4paper,journal]{IEEEtran}

\usepackage{cite}
\usepackage[utf8]{inputenc}
\usepackage{graphicx}
\usepackage{float}
\usepackage{color, colortbl}
\usepackage{xcolor}
\usepackage{array}
\usepackage{multirow}
\usepackage{footnote}
\usepackage{cite}
\usepackage{threeparttable}
\usepackage{amsmath}
\usepackage{mathtools}
\usepackage{flushend}
\usepackage[linesnumbered,vlined,norelsize]{algorithm2e}

\makeatletter
\newcommand{\removelatexerror}{\let\@latex@error\@gobble}
\makeatother

\usepackage{lipsum}

\title{Implementation and Experimental Evaluation of a Collision-Free MAC Protocol for WLANs}
      \author{\IEEEauthorblockN{Luis Sanabria-Russo\IEEEauthorrefmark{3}, Francesco Gringoli\IEEEauthorrefmark{2}, Jaume Barcelo\IEEEauthorrefmark{3}, Boris Bellalta\IEEEauthorrefmark{3}}\\
      \IEEEauthorblockA{\IEEEauthorrefmark{3}Department of Information and Communication Technologies \\ Universitat Pompeu Fabra, Barcelona, Spain
      \\\ {\tt \{luis.sanabria,jaume.barcelo,boris.bellalta\}@upf.edu}}\\
      \IEEEauthorblockA{\IEEEauthorrefmark{2}Dipartamento di Elettronica per l'Automazione \\Universit\`{a} degli Studi di Brescia, Brescia, Italy
      \\\ {\tt francesco.gringoli@ing.unibs.it}}
  }

\begin{document}

\maketitle


\begin{abstract}
Collisions are a main cause of throughput degradation in Wireless LANs. The current contention mechanism for these networks is based on a random backoff strategy to avoid collisions with other transmitters. Even though it can reduce the probability of collisions, the random backoff prevents users from achieving Collision-Free schedules, where the channel would be used more efficiently. Modifying the contention mechanism by waiting for a deterministic timer after successful transmissions, users would be able to construct a Collision-Free schedule among successful contenders. This work shows the experimental results of a Collision-Free MAC (CF-MAC) protocol for WLANs using commercial hardware and open firmware for wireless network cards which is able to support many users. Testbed results show that the proposed CF-MAC protocol leads to a better distribution of the available bandwidth among users, higher throughput and lower losses than the unmodified WLANs clients using a legacy firmware.

\end{abstract}


\section{Introduction}
Wireless Local Area Networks (WLANs) are a very well-known and broadly used technology for providing wireless access to a wired network or the Internet. As more throughput is available due to advances at the Physical layer (PHY)~\cite{6191306}, there is an ever-increasing interest on providing WiFi connectivity everywhere, ranging from conventional Small-Office/Home-Office (SOHO) environments to campuses or stadiums. 

These new scenarios carry new challenges in terms of resource allocation. Given that WiFi networks operate over a unlicensed spectrum band called the Industrial, Scientific and Medical (ISM) radio band (particularly in the 2.4-2.5 GHz and 5.725-5.875 GHz bands), packing many users into these limited bands will challenge the advertised throughput. One of the main causes of this throughput degradation derives from transmissions performed at the same time over the same WiFi channel, causing collisions among transmitting users, i.e., an unintelligible message to the intended receiver, therefore wasting channel time and thus reducing the system throughput.

Although the IEEE 802.11 standard adopts the Carrier Sense Multiple Access with Collision Avoidance (CSMA/CA) at the Medium Access Control (MAC) level to {\it avoid} collisions, it is not able to prevent them completely. CSMA/CA is implemented in WLANs through a Distributed Coordination Function (DCF) that in turn is based on a Binary Exponential Backoff (BEB) mechanism which defers each user's transmission for a random number of empty slots drawn from a Contention Window (CW). However, as the number of competing stations increases, the probability that two users start transmitting at the same time gets higher, leading to collision slots of an approximate duration equal to those slots that contain successful transmissions. 

To avoid collapsing the network, colliding nodes double their respective CW. On the contrary, when only one transmits, the successful slot embeds also a requirement for an acknowledgment frame addressed to the original transmitter, which resets the CW to the default minimum upon the reception of the said acknowledgement frame. In any case, the time wasted in collisions contribute to the degradation of the throughput.


Carrier Sense Multiple Access with Enhanced Collision Avoidance (CSMA/ECA)~\cite{research2standards} is also a totally distributed MAC protocol for WLANs. CSMA/ECA is capable of constructing a Collision-Free schedule by deferring the transmission of successful users deterministically (i.e., using a deterministic backoff after each successful transmission) instead of using a random backoff, as CSMA/CA does. This way, users that successfully transmitted in the past will schedule future transmissions without the possibility of colliding with other successful users in future cycles.

There are many studies regarding the performance of CSMA/ECA that show how the enhanced collision avoidance mechanism is capable of achieving greater throughput and with a greater number of contenders than the current MAC for WiFi~\cite{research2standards,fairness-ECA,CSMA_ECA,bellalta2009vtc,params_ECA} . Furthermore, as CSMA/ECA deviates very little from CSMA/CA its implementation in commercial hardware using an open firmware like OpenFWWF~\cite{OpenFWWF} requires little modification.

Previous experimental studies, like~\cite{ECA-DEMO-INFOCOM14, sanabria2013prototyping, BECA-test}, show that Collision-Free operation with OpenFWWF and CSMA/ECA can be achieved only for high values of the deterministic backoff. In particular, when using short deterministic backoff values, like 8, 16 or 32 slots, stations failed to maintain a collision-free operation for the length of the experiments, whether due to lack of time precision or misinterpretation of the state of the channel before transmission (caused by an imperfect Clear Channel Assessment (CCA) mechanism).


To avoid such problems a different approach is followed in this work. In order to ensure precision in the scheduling mechanism a more accurate set of instructions is implemented at firmware level. These modifications make use of a continuous timer to schedule transmissions instead of a backoff based on discrete slots. Further, possible problems with the CCA mechanism are avoided by sensing the channel for a period equivalent to only two empty slots before the scheduled transmission. 

This work shows the experimental results of a Collision-Free MAC (CF-MAC) protocol for WLANs that uses a deterministic timer after successful transmissions, similar to CSMA/ECA's deterministic backoff; nevertheless its implementation is done through more precise firmware instructions which force the nodes to attempt transmission exactly when scheduled, avoiding the inaccuracies experienced in~\cite{ECA-DEMO-INFOCOM14, sanabria2013prototyping, BECA-test}.


Results show that using a continuous timer instead of counting slots allows successful users to maintain longer collision-free schedules and to achieve a better distribution of the available bandwidth when compared to CSMA/CA.

%


The details of CSMA/ECA are described in Section~\ref{defineECA}, while our Collision-Free MAC protocol and the tools used to implement it on comercial hardware are discussed in Section~\ref{defineOpenFWWF}. The testbed is detailed in Section~\ref{defineTestbed} while the results and drawn conclusions appear in Section~\ref{defineResults} and Section~\ref{defineConclusions}, respectively.


\section{Carrier Sense Multiple Access with Enhanced Collision Avoidance}\label{defineECA}
CSMA/ECA is a totally distributed and collision-free MAC protocol for WLANs. It differs slightly from CSMA/CA in that nodes use a deterministic backoff after successful transmissions. 

Users or \emph{contenders} in a WLAN schedule transmissions based on a \emph{Backoff Counter}, $B$. At startup, in both CSMA/CA and CSMA/ECA this Backoff Counter is drawn randomly and uniformly, $B\in[0,2^{k}\text{CW}_{\min}-1]$; where CW$_{\min}$ is the minimum Contention Window of typical value CW$_{\min}=16$ and $k\in[0,m]$ is the backoff stage with initial value of $k=0$ and maximum value of $k=m=6$. 
After $B$ number of empty slots have passed, the contender will attempt transmission and wait for an \emph{acknowledgement} (ACK) from the receiver. If no ACK is received, a collision is assumed and a retransmission is scheduled.

\begin{figure*}[tb!]
\centering
  \includegraphics[width=0.8\linewidth]{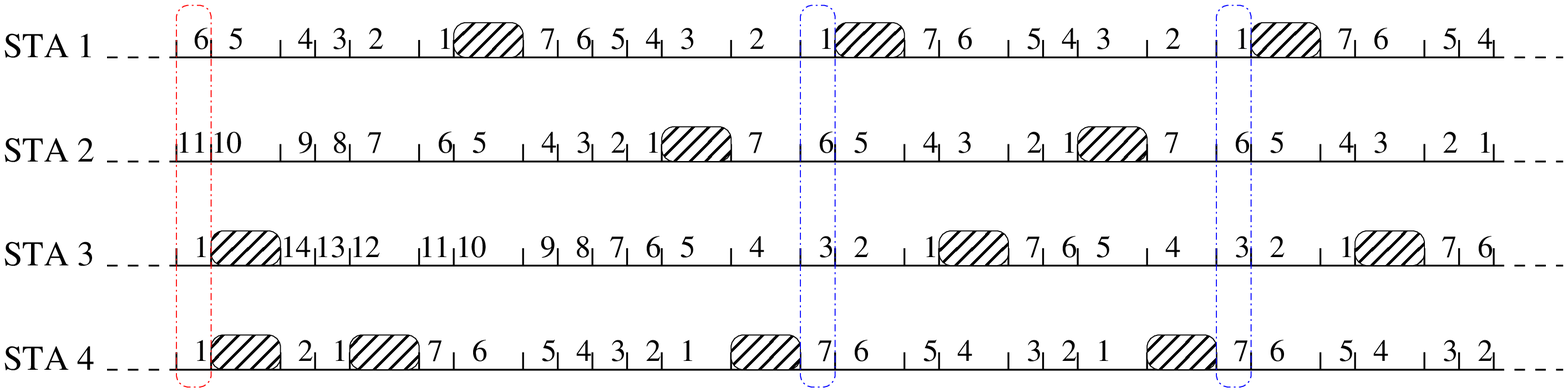}
  \caption{CSMA/ECA example in saturation: all contenders have a message to transmit all the time.}
  \label{fig:BECA-example}
\end{figure*}

For both CSMA/CA and CSMA/ECA, when a collision is detected the affected contenders will recompute their Backoff Counter increasing the backoff stage by one ($k\leftarrow k+1$). Nevertheless, the handling of a successful transmission is the main difference between these two protocols.

When an ACK is received, CSMA/CA contenders reset their backoff stage ($k\leftarrow 0$) and recompute the Backoff Counter. Whereas CSMA/ECA, after resetting the backoff stage, instructs nodes to use a deterministic backoff, $B_{d}=CW_{\min}/2$ after a successful transmission. This value of $B_{d}$ is roughly equal to the expectation of the backoff counter chosen by CSMA/CA at the initial backoff stage; thus providing fairness between CSMA/ECA and CSMA/CA stations~\cite{research2standards}.

This switch to $B_{d}$ avoids collisions among successful transmitters and thus increases the throughput for CSMA/ECA contenders. Fig.~\ref{fig:BECA-example} shows the dynamics of CSMA/ECA with four users.

\begin{figure}[tb]
\centering
  \includegraphics[width=\linewidth]{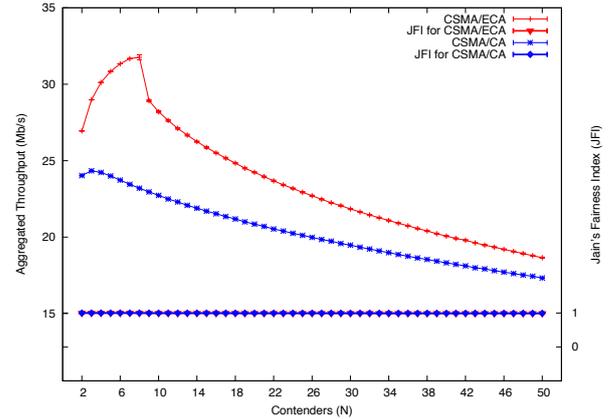}
  \caption{CSMA/ECA vs. CSMA/CA: throughput. (Using a $65$ Mb/s PHY.)}
  \label{fig:BECA-throughput}
\end{figure}

In Fig.~\ref{fig:BECA-example}, the horizontal lines represent four contenders, while the numbers are the number of empty slots until the expiration of their respective Backoff Counters. The first outline indicates that STA-3 and STA-4 collide because their Backoff Counter is the same, $B=1$. After the collision, both stations recompute their backoff; for STA-3: $B=14$, and for STA-4: $B=2$. Having passed two empty slots STA-4 transmits and picks a deterministic backoff, $B_{d}=7$ in this case. When all stations have transmitted successfully, a collision-free schedule is built. Fig.~\ref{fig:BECA-throughput} (redrawn from~\cite{research2standards}) shows the achieved throughput and the Jain's Fairness Index (JFI)~\cite{JFI} for both CSMA/ECA and CSMA/CA obtained from computer simulations.

The collision-free schedule that is built with CSMA/ECA is responsible for the increase in the aggregated throughput shown in Fig.~\ref{fig:BECA-throughput}. Further, the value of $JFI = 1$ suggests an even distribution of the available throughput regardless of the number of contenders ($N$). Nevertheless, when $N$ surpasses the value of the deterministic backoff ($B_{d}$), collisions reappear. This effect degrades CSMA/ECA throughput and approximates it to CSMA/CA's.

Further enhancements like \emph{Hysteresis} and \emph{Fair Share} presented in~\cite{research2standards} provide an increase in the number of contenders CSMA/ECA can accommodate in a collision-free schedule; later called CSMA/ECA$_{\text{Hys+FS}}$ to distinguish it from basic CSMA/ECA.

\subsection*{CSMA/ECA prototypes in real hardware}
One of the main advantages of CSMA/ECA in terms of implementation is that it does not deviate too much from the current MAC. This allows the use of open firmware that already contains the base code for CSMA/CA to be modified towards CSMA/ECA.

The CSMA/ECA implementations~\cite{ECA-DEMO-INFOCOM14, sanabria2013prototyping, BECA-test} manage to change the backoff mechanism after a successful transmission. Nevertheless, the CSMA/ECA behaviour was maintained only for a short time: it turned out that contenders were not able to accurately keep the deterministic schedule. This effect can be the result of the cards' imperfections while decrementing the counter when the channel is sensed idle, or failing to freeze it when another node's transmission is taking place.



In this work we implement a Collision-Free MAC protocol for WLANs using a deterministic timer that instructs nodes precisely when to attempt transmission, without being frozen by ongoing transmissions in the channel (contrary to CSMA/CA's backoff counter). Furthermore, to avoid possible CCA imperfections from causing a disruption of a collision-free schedule, the carrier sense algorithm is activated only two slots before each transmission attempt, mostly to avoid collisions with ongoing transmissions. A similar approach was proposed for ALOHA-based wireless multi-hop networks~\cite{jaume-mesh} to achieve collision-free operation. The approach of~\cite{jaume-mesh} is evaluated only analytically and by simulation.


\section{The Collision-Free MAC (CF-MAC)}\label{defineOpenFWWF}
Several manufacturers embraced the Soft-MAC~\cite{neufeld2005softmac} approach for
interconnecting their Wi-Fi Network Interface Cards (NICs) with
general purpose systems. A dedicated CPU on the NIC controls the radio
circuitry and pulls complete 802.11 frames prepared by the main Operating System (OS) kernel from
an interconnecting bus, e.g., a PCI bus, and schedules their
transmission in real time. Thanks to this approach, the NIC offloads
the time-critical actions related to the channel access, while the
main kernel controls all other functionalities. The
CPU on the NIC runs the MAC algorithm by executing a software (the firmware from here on) that reacts to transmission/reception
history and drives the evolution of the Contention Window and
the Backoff Counter. By replacing the firmware, one can
deeply customize the MAC or even switch to a different one, e.g.,
Time Division Multiple Access (TDMA)~\cite{WMP}, instead of CSMA/CA.

To build and test our Collision-Free MAC protocol we chose the Open FirmWare for Wi-Fi
networks (OpenFWWF~\cite{OpenFWWF}) as it is the only open source
firmware ever released for controlling Wi-Fi NICs. Specifically, it is
compatible with the Airforce54~\cite{broadcom-airforce} chipset family
from Broadcom. Together with the \texttt{b43} Linux
driver~\cite{b43-info} it already showed up as a flexible research
platform~\cite{gringolitmc14,gringoliccr14}: given the low
per-node price\footnote{A Linksys WRT54GL node includes a MIPS main
  CPU and a Broadcom 4318 NIC and was quoted as low as $\$39$} it
also allows inexpensive deployment of dense Linux based testbeds.

For this work we built on the CSMA/CA code available in OpenFWWF and added a new packet scheduling mechanism as we detail in the following.

\subsection{Protocol Description}

\begin{figure*}[tb]
\centering
  \includegraphics[width=0.8\linewidth]{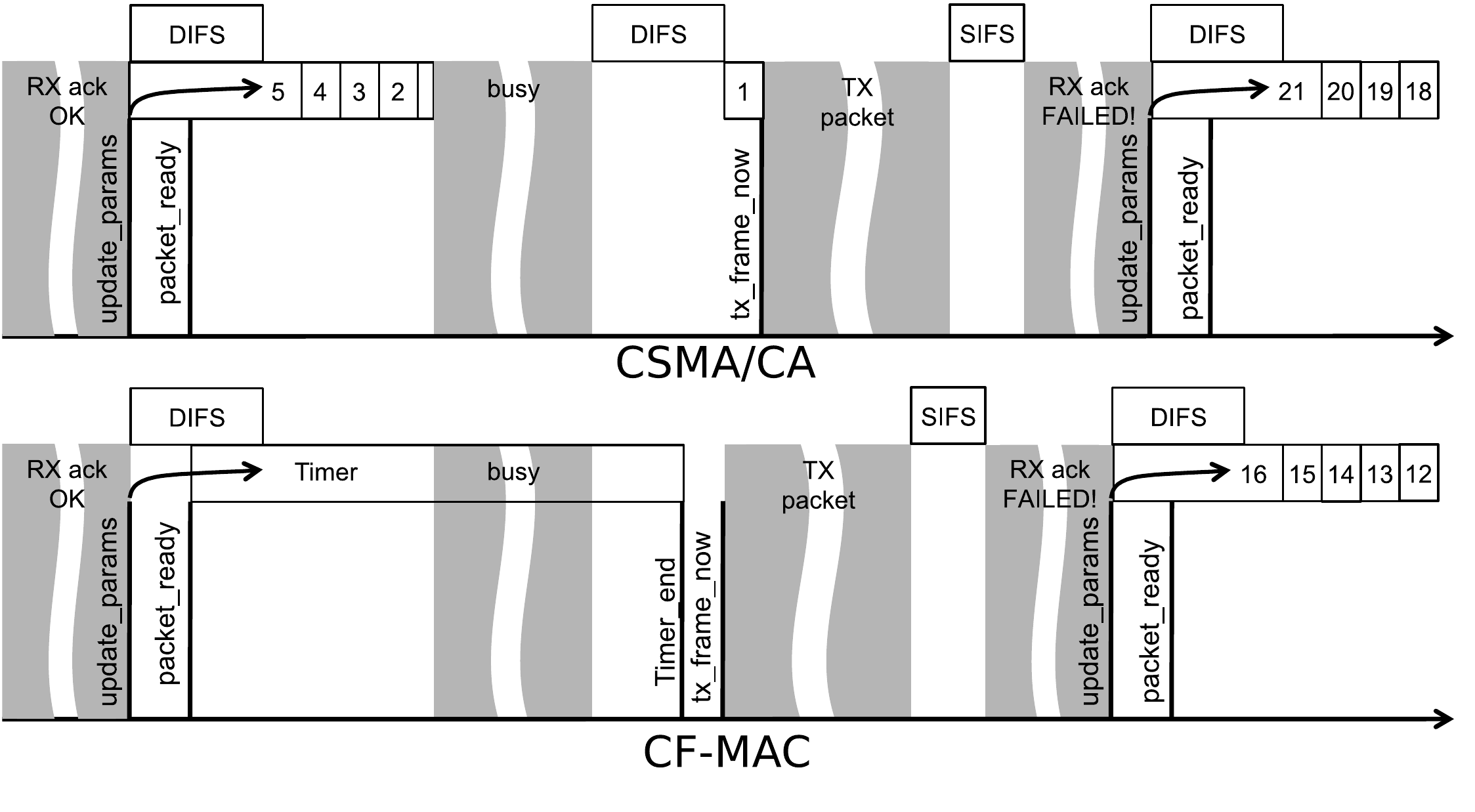}
  \caption{CSMA/CA (up) and CF-MAC (down) channel access with success and failure}
  \label{fig:accesses}
\end{figure*}

OpenFWWF implements a simple State Machine (SM) for controlling the
hardware in real time. The SM evolution is driven by a main loop that
reacts to events by executing specific handlers\footnote{In the
  following we consider only the limited subset of events that were
  changed to implement CF-MAC.}. When a packet, originally prepared
by the Linux kernel, is ready in the NIC memory, handler {\tt
  packet\_ready} sets up the radio hardware according to the packet
meta data (e.g., it fixes rate, modulation format, and power level),
schedules the transmission and jumps back to the main loop. Then, the
Transmission Engine (TXE) takes care of accessing the channel,
i.e., it decrements the Backoff Counter ($B$) according to the
Distributed Channel Function (DCF) rules and it eventually starts the
actual transmission. This triggers the execution of the~{\tt
  tx\_frame\_now}~event that prepares the ACK time-out clock and
finalizes the MAC header\footnote{As transmission has already
  started, these actions must be completed before the physical preamble
  end.}. If the ACK-frame is received or if the ACK time-out expires
and the maximum number of attempts for this packet is reached, handler
{\tt update\_params} resets the Contention Window to the minimum (CW$_{\min}$),
otherwise it doubles the CW. Finally it loads the $B$ counter with a
fresh value.

As the TXE stops counting down $B$ when it detects the channel busy, we could not use it for implementing our Collision-Free MAC protocol, i.e., by loading the $B$ counter with a value proportional to the exact schedule delay. Further, as experimented in~\cite{ECA-DEMO-INFOCOM14, sanabria2013prototyping, BECA-test} nodes would quickly go out-of-sync. To avoid these unpredictable backoff inaccuracies we exploited another feature as in~\cite{berger14} that
allows the firmware to start the immediate transmission of a frame,
independently of the channel conditions (we call this feature TX$_{\text{now}}$ from this point onwards). We hence reworked the main loop to make it continuously check if the schedule delay has elapsed. In this case it invokes the TX$_{\text{now}}$ code if the channel is found idle, otherwise it keeps checking the channel for a period equivalent to a couple of empty slots to better avoid collisions with ongoing transmissions, e.g., tails of previous frames in the global schedule. If the channel is still busy, it will backoff for a period equivalent to a random number of slots drawn from a Reduced Window (RW$=7$). In the case that the channel is found busy again, the node aborts this round and goes back to CSMA/CA.


\subsection{Implementation}

The modifications required to make use of the TX$_{\text{now}}$ feature have the following effects:

	\begin{enumerate}
		\item It supposes a modified use of the carrier sense algorithm. That is, the node only listens to the channel for a short period equivalent to two empty slots before attempting the transmission.
		\item For using the TX$_{\text{now}}$ instruction a different approach should be used. Therefore, we create a timer based on real time (measured in $\mu$s).
		\item The use of a timer ($T_{\text{c}}(N,r)$) is subject to the number of contenders ($N$) and rate ($r$).

%
		\end{enumerate}

$T_{\text{c}}(N,r)$ basically is the duration of all transmissions in a schedule with $N$ contenders at a certain rate, plus the duration of the reception of the ACK for such transmissions and a short guard interval ($\epsilon$). 

				
Notice that $T_{\text{c}}(N,r)$ supposes a previous knowledge of the number of contenders, making the obtained results useful for research purposes but with little practical use in real world WLANs. Table~\ref{tab:timer-values} shows the value of $T_{\text{c}}(N,r)/N$ for each rate we tested our Collision-Free MAC protocol. It is important to highlight that the value of $\epsilon$ was adjusted in order to find a value of $T_{\text{c}}(N,r)$ capable of accommodating all $N$ contenders. This adjustment is based on experimental results.

		\begin{table}[tb]
			\centering
			\caption{\ensuremath{T_{\text{c}}(N,r)/N} and \ensuremath{\epsilon} values for different rates}
			\label{tab:timer-values}
			\begin{tabular}{|c|c|c|c|}
				\hline
				$r$ (Mb/s) & $T_{\text{c}}(N,r)/N$~($\mu$s)& $\epsilon$~($\mu$s) & $T_{\text{c}}(N,r)/N+\epsilon$\\
				\hline
				$6$ & $2233.5$ & $91.5$ & $2325$\\
				$11$ & $1567.5$ & $132.5$ & $1700$\\
				$12$ & $1197.5$ & $102.5$ & $1300$\\
				$24$ & $681.5$ & $106.5$ & $788$\\
				$48$ & $421.5$ & $103.5$ & $525$\\
				\hline
			\end{tabular}
		\end{table}

Algorithm~\ref{alg:BECA-Tx-now} shows an example of the proposed protocol. Basically, stations substitute the random backoff $B$ by the $T_{\text{c}}(N,r)$ timer after a successful transmission. Successful nodes will continue to attempt transmission every $T_{\text{c}}(N,r)~\mu$s, until two consecutive collisions are detected or the channel is found busy for too long, after which a random backoff is drawn and the node goes back to CSMA/CA operation (this process is detailed at line~\ref{deterministicBackoff-Tx}). Allowing successful nodes to \emph{stick} with the deterministic timer even after collisions was first proposed in~\cite{L_MAC2} and called \emph{stickiness} (also tested with CSMA/ECA~\cite{jaume2011towards}). This allows contenders to converge faster towards a collision-free schedule. Further, when collision-free operation is achieved, it prevents successful users from going back to a random operation due to channel errors.

\begin{algorithm}[tb]\caption{\small{Overview of the packet scheduling mechanism for CF-MAC.}}\label{alg:BECA-Tx-now}
	\While{the device is on}
	{
  		$ret \leftarrow 0$ ; $k \leftarrow 0$\;
  		$B \leftarrow \mathcal{U}[0,2^k{\rm{CW}_{min}}-1]$\;
  		\While{there is a packet to transmit}{
    			\Repeat{($ret = R$) or (success)}{
      				\While{$B>0$}{
        					wait 1 slot\;
        					$B \leftarrow B-1$\;
      				}
      				\colorbox{yellow}{Attempt transmission of 1 packet;}\\
      				\If{collision}
      				{
        					$ret \leftarrow ret+1$\;
        					$k \leftarrow \min (k+1,m)$\;
        					$B \leftarrow \mathcal{U}[0, 2^k {\rm{CW}_{min}} -1]$\;
      				}
    			}
    			$r \leftarrow 0$\;
    			$k \leftarrow 0$;\\
    			\eIf{$ret = R$}
	 		{
      				Discard packet\;
	  		}
			{\Repeat{(two consec. collisions {\bfseries or} busy two times)}{
				wait $T_{\text{c}}(N,r)$ seconds\;\label{deterministicBackoff-Tx}
				\colorbox{yellow}{Attempt transmission of 1 packet;}\\
			}}
			$B \leftarrow \mathcal{U}[0, 2^k {\rm{CW}_{min}} -1]$\;
   		}
  		Wait until there is a packet to transmit\;
	}
\end{algorithm}

\section{Testbed description}\label{defineTestbed}
\begin{figure}[tb]
\centering
  \includegraphics[width=0.9\linewidth]{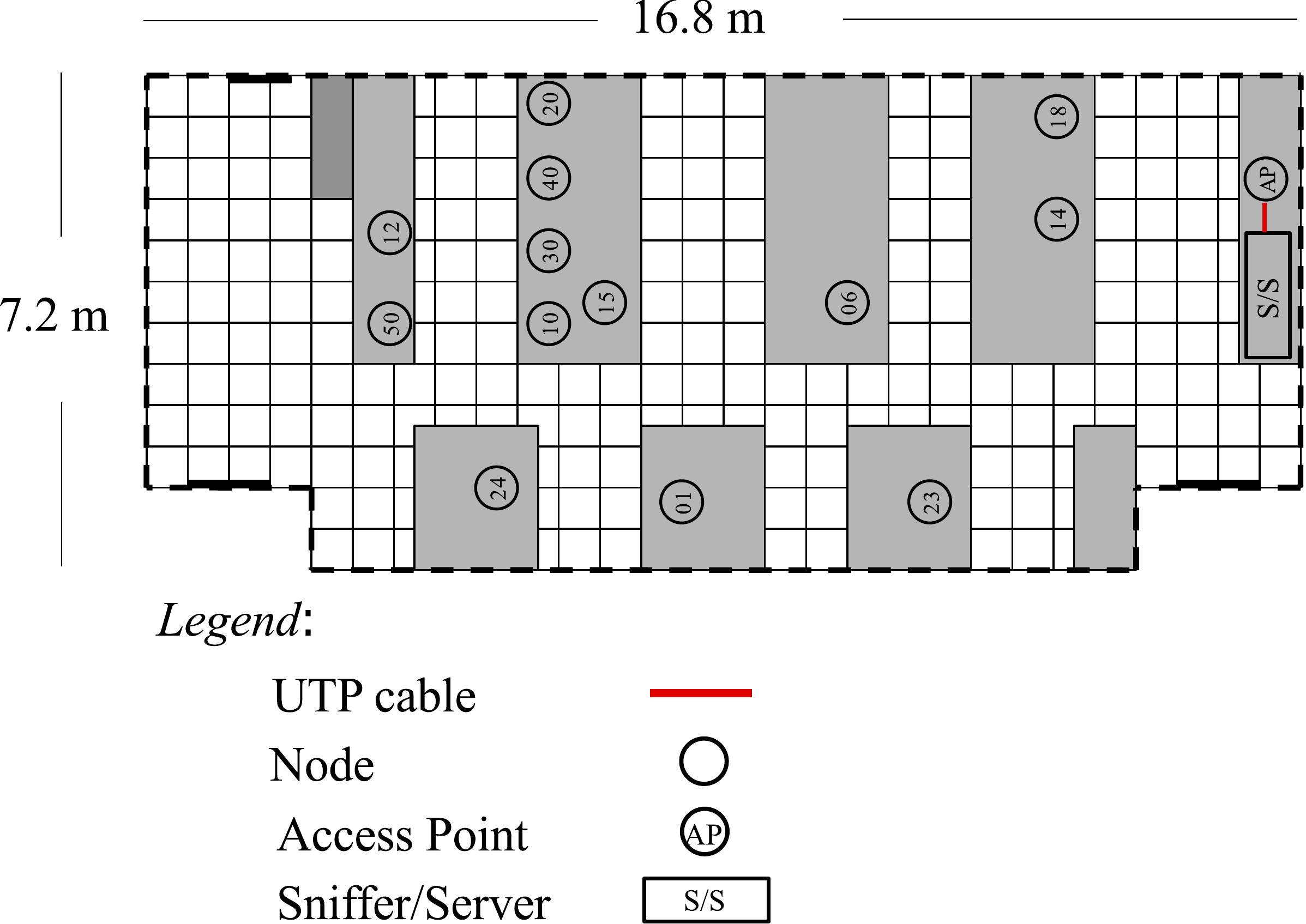}
  \caption{Testbed.}
  \label{fig:testbed}
\end{figure}

Each node used for the testing of our prototype is equipped with a commercial WiFi card compatible with both OpenFWWF and the \texttt{b43} driver. The modified firmware was loaded into the twelve testing nodes which were arranged mimicking a conventional workspace environment: placed at different distances from an AP and using a free WiFi channel in order to avoid external interferences from other networks. Fig.~\ref{fig:testbed} is a graphic representation of the nodes' layout, while Table~\ref{tab:tesbed-specs} gathers the PHY and MAC settings used.

\begin{table}[tb]
	\centering
	\caption{PHY and MAC parameters for the testbed}
	\label{tab:tesbed-specs}
	\begin{tabular}{|c|c|}
		\hline
		\multicolumn{2}{|c|}{{\bfseries PHY}}\\
		\hline
		{\bfseries Parameter} & {\bfseries Value}\\
		\hline
		PHY rate~(Mb/s) & $6,11,12,24,48$\\
		Empty slot~($\mu s$) & $9$ ($20$ for $11$ Mb/s)\\
		DIFS~($\mu s$) & $28$ ($50$ for $11$ Mb/s)\\
		SIFS~($\mu s$) & $10$\\
		\hline
		\multicolumn{2}{|c|}{{\bfseries MAC}}\\
		\hline
		{\bfseries Parameter} & {\bfseries Value}\\
		\hline
		Maximum backoff stage ($m$) & 6\\
		Minimum Contention Window ($CW_{\min}$) & 16\\
		Maximum retransmission attempts & 6\\
		Packet size (Bytes) & 1470\\
		Duration of each test ($s$) & 90\\
		\hline
	\end{tabular}
\end{table}

Upon each test, the rate of each station is fixed and a unidirectional iPerf~\cite{tirumala2005iperf} session is established from each station to a Server using UDP. Each transmitter then is saturated (is always attempting to transmit) for a period of ninety seconds. We used also another node, a Sniffer, to capture all traffic using TCPdump~\cite{jacobson2003tcpdump}. The Server and the Sniffer are represented in Fig.~\ref{fig:testbed} as a single station, connected via Ethernet to the Access Point (AP).

To better analyse the behaviour of our protocol we added to the stations' firmware a counter for the number of successful transmission, i.e., those acknowledged by the receiver; and a counter for the number of failures, i.e., those not acknowledged. It is possible to derive several metrics by analysing the captured traces and the counters, like:

\begin{itemize}
	\item {\bfseries Throughput per station:} by looking at the log of each iPerf session, it is possible to obtain an estimation of the achieved throughput of each station. Further, by looking at the number of successfully sent packets (counted by the firmware) a measure of throughput can also be derived.
	\item {\bfseries Inter-arrival time:} is the time between the transmission of two frames by the same station. This metric reflects the time invested in the contention mechanism.
	\item {\bfseries Fraction of lost frames:} each time a station attempts a retransmission, the result of the previous transmission is counted as a failure. Knowing the number of failures and the total number of transmission attempts, the fraction of lost frames is computed.
	
\end{itemize}

In the following section we show and discuss the results of the experiments that we performed. In each experiment we used nodes configured in the same way, e.g., they were all using our CF-MAC protocol, or CSMA/CA.

\section{Results}\label{defineResults}

\subsection{Throughput and Fairness}
When comparing both protocols it is useful to look at the achieved throughput, but also at how the available bandwidth is distributed among the contenders.

\begin{figure}[tb]
\centering
  \includegraphics[width=0.95\linewidth]{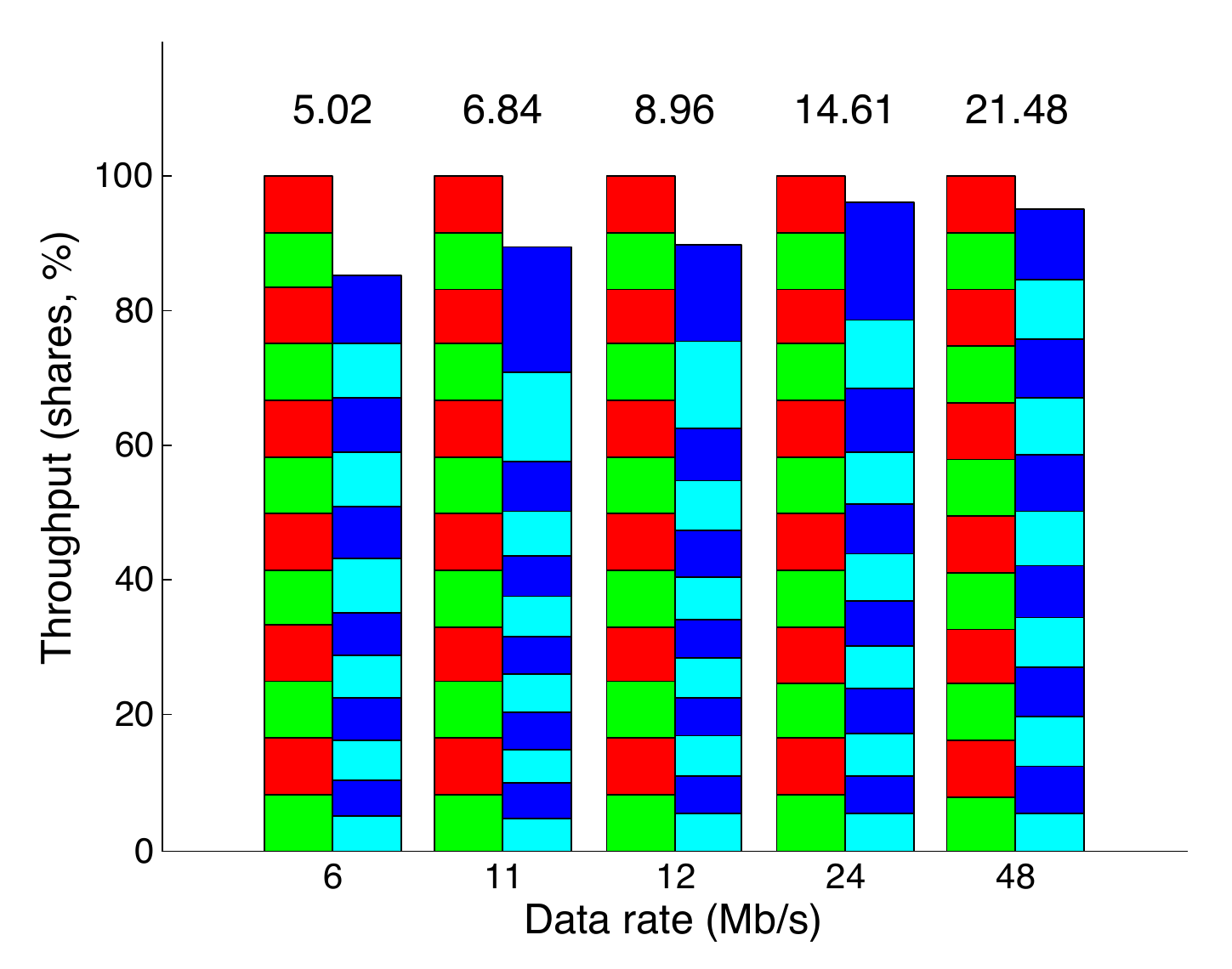}
  \caption{Throughput for different data rates. Each experiment if conformed by twelve nodes. The left bar represents our CF-MAC and the right one CSMA/CA. Hovering numbers indicate the accumulated throughput achieved by CF-MAC in Mb/s.}
  \label{fig:throughput}
\end{figure}

\begin{figure}[tb]
\centering
  \includegraphics[width=0.95\linewidth]{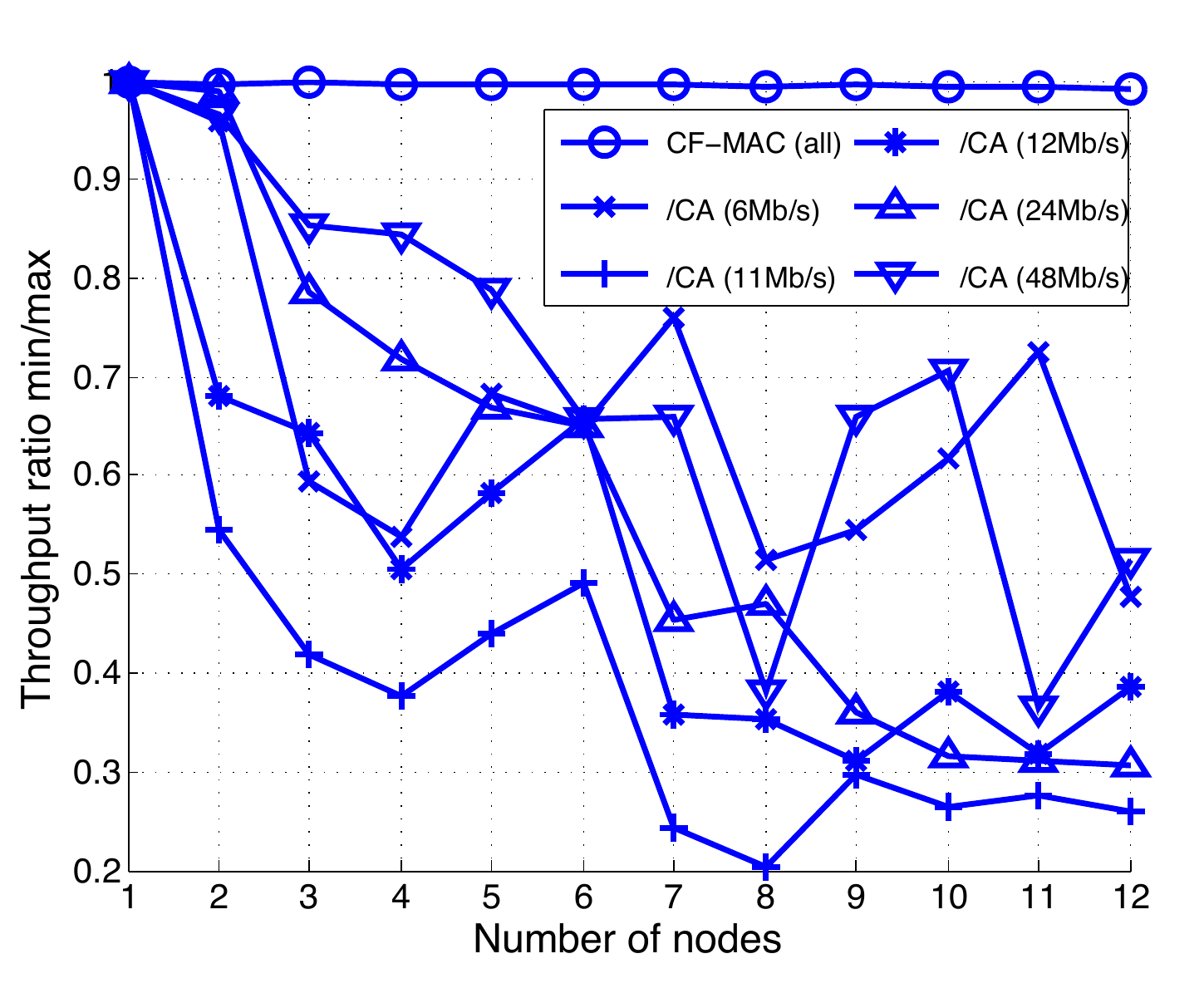}
  \caption{Min/Max throughput ratio. /CA curves represent CSMA/CA experiments.}
  \label{fig:fairness}
\end{figure}

Fig.~\ref{fig:throughput} shows ten independent experiments with increasing data rates and twelve nodes each. For each x-point, two separate experiments are shown, one for a network composed of only CSMA/CA nodes (right bar) and one for nodes loaded with our CF-MAC protocol (left bar). Each bar is divided in boxes which represent the throughput share of a station.

Given that CF-MAC stations are able to construct a Collision-Free schedule using the $T_{\text{c}}(N,r)$ timer after a successful transmission, the channel is used more efficiently. Whereas CSMA/CA stations waste time recovering from collisions and in contention for the channel.

We can see that the CSMA/CA network achieves less cumulative throughput than CF-MAC for all the tests performed. Further, the throughput is not evenly distributed among the contenders. Boxes in each bar represent the throughput of the corresponding node that we sorted decreasingly top to bottom: we can clearly see that in the case of CSMA/CA the top boxes are much taller than the bottom ones, while for the CF-MAC protocol the boxes are equally shaped. This effect is underlined in Fig.~\ref{fig:fairness} that shows the min/max throughput ratio for network setups with increasing number of nodes and for all the experiments (different rates) that we performed. CF-MAC with any number of nodes and all tested rates shows that the throughput is efficiently shared among contenders, whereas different CSMA/CA network setups (denoted as /CA in Fig.~\ref{fig:fairness}) show an uneven distribution of the available throughput.

\subsection{Inter-arrival Times}
\begin{figure}[tb]
\centering
  \includegraphics[width=0.95\linewidth]{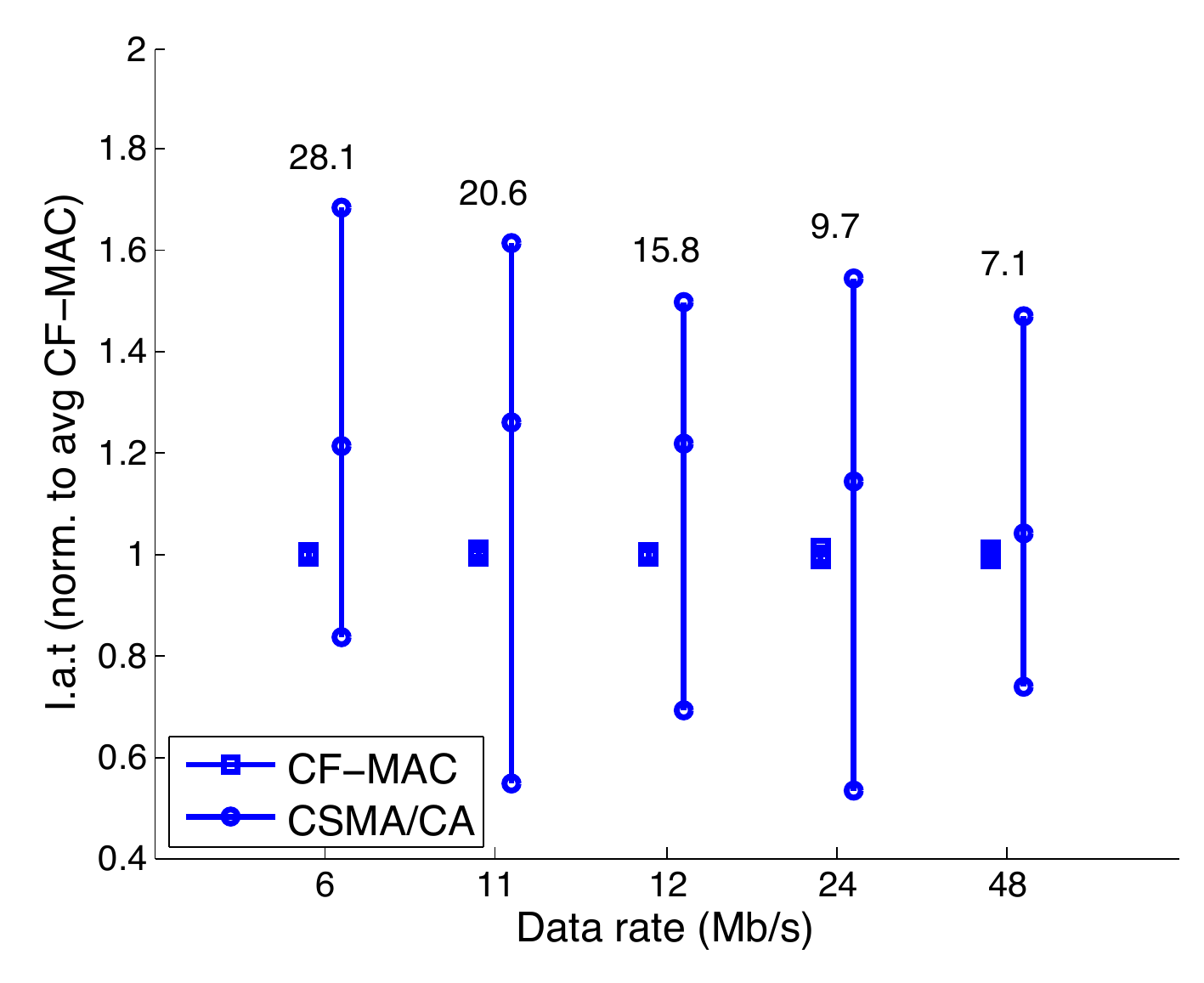}
  \caption{Inter-arrival Times (normalized to the average of CF-MAC). For each X-axis point: left boxes represent CF-MAC stations, while the right circles are CSMA/CA users. Middle circles represent the average among all CSMA/CA nodes, while higher and lower circles represent the maximum and minimum respectively. Hovering numbers are the average I.a.t for CF-MAC.}
  \label{fig:iat}
\end{figure}

CSMA/CA nodes pick $B$ randomly and freeze it when a transmission is being performed in the channel, which translates in a variable inter-arrival time; while CF-MAC stations schedule transmissions according to the predefined timer ($T_{\text{c}}(N,r)$). This is made evident by Fig.~\ref{fig:iat}, where the time between consecutive transmissions varies considerably more for CSMA/CA than for CF-MAC. This suggests that CSMA/CA nodes on average spend more time in contention and recovering from collisions, also contributing to the throughput degradation.

\subsection{Lost Frames}
\begin{figure}[tb]
\centering
  \includegraphics[width=0.95\linewidth]{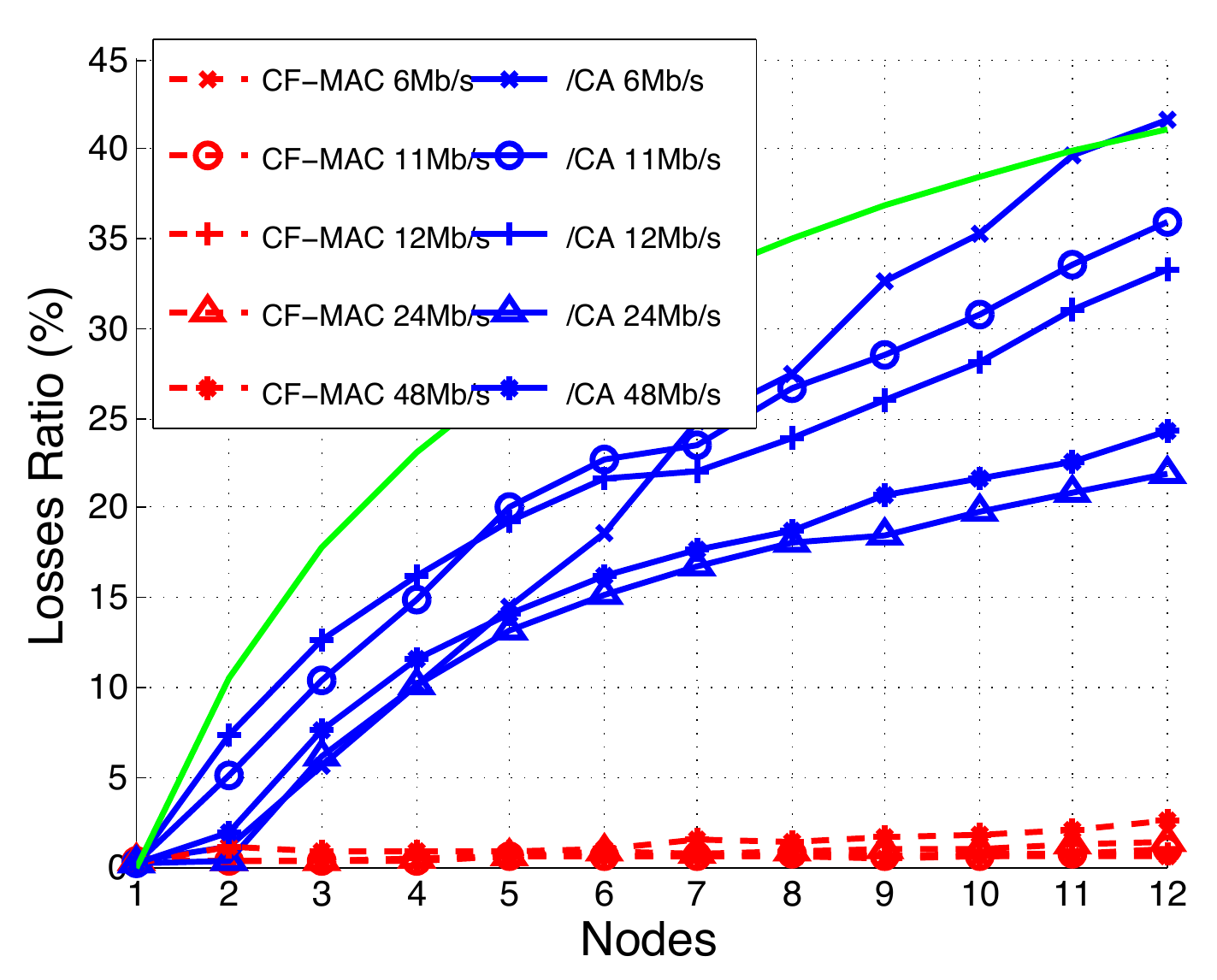}
  \caption{Fraction of losses.}
  \label{fig:losses}
\end{figure}

To derive a measure of the average losses per node we used the values of the two counters added to the firmware: we counted the number of failed transmissions $F$ (indicated by the lack of reception of an ACK), successful transmissions $S$ (when an ACK is received) and the computed the number of transmission attempts: $A=F+S$. Fig.~\ref{fig:losses} shows the losses ratio ($F/A$) for CSMA/CA and CF-MAC alongside a reference curve derived from the model proposed in~\cite{bianchi2000performance}.

CSMA/CA stations suffer from an increased number of collisions, mostly due to the randomness of the backoff mechanism; whereas CF-MAC nodes enjoy a much reduced number of collisions due to the implementation of the deterministic timer, $T_{\text{c}}(N,r)$, after successful transmissions.

In Fig.~\ref{fig:losses}, at higher rates (24, 48 Mb/s) the losses ratio for CSMA/CA seem to be reduced with respect to the reference curve. This effect can be caused by a defective CCA mechanism on the cards. Transmissions at these rates are shorter, so transmitters are less prone to make erroneous inferences about the channel state. On the other hand, stations at lower rates should listen to the channel for longer periods of time before attempting transmission, thus increasing the probability of a misinterpretation of the channel state.


\section{Conclusions}\label{defineConclusions}
CF-MAC is able to construct Collision-Free schedules by means of using a deterministic timer after successful transmissions, which allows a better use of the available channel time in WLANs. 

In this work it is shown how using a precise schedule for transmissions allows CF-MAC to greatly reduce the fraction of collisions in comparison with CSMA/CA. Further, this reduction of wasted channel time recovering from collisions or spent in contention is reflected in a better distribution of the available bandwidth among contenders. Moreover, by using a deterministic timer after successful transmissions stations greatly reduce the variability of time between transmissions attempts, making it a suitable technique for delay-sensitive communications.

CF-MAC was tested in a real testbed, using off-the-shelf hardware and a modified firmware for the wireless cards. Many real-world un-ideal conditions, like the performance of the CCA implementation or impractical assumptions like previous knowledge of the number of contenders for setting the deterministic timer prevent this specific implementation from being an adecuate MAC protocol for WLANs. 

Nevertheless, in itself this real-world factors are enough motivation for keep attempting to unveil and enhance the internal packet scheduling mechanisms of Collision-Free MAC protocols using deterministic backoffs in WLANs.

\section*{ACKNOWLEDGMENT}
This work was partially supported by the Spanish government, through the project CISNETS (TEC2012-32354).

\bibliographystyle{IEEEtran}
\bibliography{ref}

\end{document}